\documentclass [11pt] {article}      
\begin{document}

\def\aln{{\left ( {\alpha \over n} \right )}}
\def\ds{\displaystyle}
\def\xp{x^\prime}
\def\rb{{\bf r}}
\def\pb{{\bf p}}
\def\pp{{p^{\prime}}}
\def\pbp{{{\bf p}^{\prime}}}
\def\bra{{\langle}}
\def\ket{{\rangle}}
\def\ep{{\varepsilon}}

\def\tp{t^\prime}
\def\di{\partial}
\def\ni{\noindent}
\def\phib{\overline \phi}
\def\psib{\overline \psi}
\def\oneha{{1\over 2}}
\def\vss{\vskip .5cm}

\vss \vss 

\title{ A variational calculation of particle-antiparticle bound states 
in the scalar Yukawa model}
\author{Bingfeng Ding and Jurij Darewych\\
Department of Physics and Astronomy\\
York University\\
 Toronto, Ontario\\
M3J 1P3, Canada}
\maketitle


\begin{abstract}
We consider particle-antiparticle bound states in the scalar Yukawa 
(Wick-Cutkosky) model. The variational method in the Hamiltonian formalism 
of quantum field theory is employed. A reformulation of the model is studied, in which 
covariant Green's functions are used to solve for the mediating field in terms of the 
particle fields. A simple Fock-state variational ansatz is used to derive a relativistic 
 equation for the particle-antiparticle  states. This equation contains one-quantum-exchange
 and virtual-annihilation interactions.
   It is shown that analytic solutions of this equation can 
   be obtained for the simplified case where only the virtual annihilation interaction is retained. 
More generally, numerical and perturbative solutions of the 
equation are obtained for the massive and  massless-exchange cases.  We compare our 
results with various Bethe-Salpeter-based calculations.
\end{abstract}

\vfill\eject


{\ni \textbf {1. Introduction}}
\vskip .5truecm
The variational method has been used sparingly in treating few-particle
 bound and quasi-bound states 
in quantum field theory. Yet, it is in principle appealing, particularly
 for strongly coupled systems, because of its largely analytic and 
 non-perturbative nature. The early papers that kindled interest in the variational
 method in QFT 
are those of Schiff [1],  Coleman and Weinberg [2,3], Jackiw [4] Barnes 
and Ghandour [5], Stevenson [6], Tarrach [7] and others [8 ]. 
A brief review of this approach to few-body bound states in the Hamiltonian 
formalism of QFT up to 1996 is given in ref. [9]. 

The variational method is, of course, only as good as the trial states that are 
being employed. In the functional formulation (e.g. [5,8]) one is restricted largely 
to wave-functionals of the Gaussian type, due to the difficulty of handling 
analytically non-Gaussian functional integrals [10]. Another approach is to 
expand the 
trial state in a Fock-space basis. Indeed, the early work of Tamm [11] and 
Dancoff [12] is along these lines, though it was not variationally formulated.

The Hamiltonian formalism of QFT can be expressed in terms of the QFTheoretic 
 eigenvalue equation
\begin{equation}
{\hat H} |\Psi \rangle  = E |\Psi \rangle ,
\end{equation}
where ${\hat H}$ is the QFTheoretic 
hamiltonian operator, and $E$ is the energy  of 
the system under study. Such an equation is generally impossible to solve, except for 
some special models, particularly in 1+1 (one spatial coordinate plus time), such as
 the Thirring [13 ]and Schwinger [14] models. In the variational approach within 
the Hamiltonian formalism one seeks approximate solutions to Eq. (1) by using the 
variational principle
\begin{equation}
\delta \langle \Psi_t | {\hat H} - E | \Psi_t \rangle = 0,
\end{equation}
where $|\Psi_t \rangle$ is a suitably chosen trial state containing adjustable features 
(parameters, functions). Some examples of the application of this method to
 bound states in scalar  $\phi^6$ theory are given in refs. [6c] and [15].

One of the difficulties of the Fock-state expansion approach for realistic models 
in 3+1 (such as the Yukawa model, QED, etc.) is that it results  in an infinite 
system of coupled, multi-dimensional integral equations to be solved - an impossible 
task. Truncation, or other approximation schemes, undermine the strict variational 
nature of the approximation.  This, then, puts into question the validity of results at 
strong coupling, precisely in the domain that one wishes to address in a 
 non-perturbative approach. 
It has been pointed out recently [16] that various models can be reformulated in 
such a way that this difficulty can be circumvented.

 In this paper we shall implement  the approach given in reference [16]
 to  particle-antiparticle bound states in the  scalar Yukawa model,
in which scalar particles interact via a  mediating real (massive or massless) 
scalar field.  The treatment of two-body bound states in this model by means 
of the Bethe-Salpeter equation in the ladder approximation is known as the 
 Wick-Cutkosky model [17,18].
In addition to these original solutions of Wick and Cutkosky, 
the scalar Yukawa model has been used as a prototype QFT in many  studies. 
It has been investigated  quite extensively in various formalisms, such as the 
light-cone formulation [19-22],  various Bethe-Salpeter-based approaches 
[23-27], and others [28, 29, 30].  The 
work of Nieuwenhuis and Tjon [27b], in particular, gives a comparison of a number 
of quasipotential approximations. This makes the model appealing as a relatively simple test case.
(We make no effort here to give an exhaustive survey of the literature on this 
model.  Many relevant papers are cited in the references which we quote. A review of 
the Wick-Cutkosky model to 1988 is given by Nakanishi [31], while many mathematical 
details are given in the work of Silagadze [32].)

In the scalar yukawa model massive scalar particles interact via a  
mediating real scalar field. It is based on the Lagrangian 
density ($\hbar = c = 1$)
\begin{eqnarray}
{\cal L} &=& \di^\nu \phi^*(x)\, \di_\nu \phi (x) - 
m^2 \phi^*(x)  \phi (x)  \\    
&+& \oneha \di^\nu \chi (x) \,\di_\nu \chi (x) - \oneha 
\mu^2 \chi^2 (x)  - g \phi^*(x) \phi (x)\, \chi (x) \nonumber\    
\end{eqnarray}
The mediating ``chion'' field can be massive ($\mu \ne 0$) or massless (i.e. $\mu = 0$, as in the 
original Wick,  Cutkosky work [17, 18]). The coupling constant $g$ has dimensions 
of (mass)$^{{5 \over 2}- {N \over 2}}$ in N+1 dimensionless.  A slightly simpler model, in which 
$\phi$ is real, is often considered. In that case there are only particles and no antiparticles. 
These models are closely related, since the forces among particles (and/or antiparticles) are 
only attractive (i.e. gravity-like, rather than electromagnetic-like). Of course, in the case where
 $\phi$ is real there is no particle-antiparticle annihilation.
\vss


\ni {\bf 2. Reformulation of the model}
\vss

The fields $\phi$ and $\chi$ of the model (3) satisfy the Euler-Lagrange equations    
\begin{equation}
\di^\nu \di_\nu \chi (x)+ \mu^2 \chi (x) = -g  \phi^*(x) \phi (x),
\end{equation}
\begin{equation}
\di^\nu \di_\nu \phi (x) + m^2 \phi (x) = -g  \phi (x) \chi (x),
\end{equation}
and the conjugate of (5). Equation (4) has the formal solution
\begin{equation}
\chi(x) = \chi_0(x) + \int d\xp \,D(x-\xp)\, \rho(\xp),
\end{equation} 
where $dx = d^Nx \,dt$ in $N+1$ dimensions,  $\rho(x) = - g 
\phi^*(x) \phi (x)$, $\chi_0 (x)$ satisfies the homogeneous 
(or free field) equation (eq. (4) with $g = 0$), while $D(x-\xp)$ 
is a covariant Green function (or chion  propagator, in QFTheoretic 
language), such that
\begin{equation}
\left ( \di^\nu \di_\nu  + \mu^2 \right ) D(x-\xp) = \delta^{N+1}(x-\xp).
\end{equation}
Equation (7) does not specify $D(x-\xp)$ uniquely since, for 
example, any solution of the homogeneous equation can be added 
to it without invalidating (7). Boundary conditions based on 
physical considerations are used to pin down the form of $D$.

Substitution of the formal solution (6) into eq. (5) yields the equation
\begin{equation}
\di^\nu \di_\nu \phi(x) + m^2 \phi(x) = -g  \phi (x) 
\chi_0 (x) - g \phi(x) \int d\xp D(x-\xp) \rho(\xp).
 \end{equation}
Equation (8) is derivable from the action principle $\ds \delta 
\int dx\, {\cal L} = 0$, corresponding to the Lagrangian density
\begin{eqnarray}
{\cal L}&=&\di^\nu \phi^*(x) \di_\nu \phi (x) - m^2 \phi^*(x)  
\phi (x) 
  - g \phi^*(x) \phi (x)\, \chi_0 (x) \\
&+& \oneha \int d\xp \rho (x) D(x-\xp) \rho(\xp),  \nonumber\
\end{eqnarray}
provided that $D(x-\xp) = D(\xp-x)$. (We suppress the free chion part of the Lagrangian.)

The QFTs based on (3) and (9) are equivalent in that, in conventional 
covariant perturbation theory, they lead to the same invariant 
matrix elements in various order of perturbation theory.  
The difference is that, in the formulation based on (9), the 
interaction term that contains the propagator leads to Feynman 
diagrams  involving virtual chions, while the  
term that contains $\chi_0$  correspond to diagrams that cannot be generated 
using the term with $D(x-\xp)$, such as those with external (physical) chion lines.

The Hamiltonian density corresponding to the Lagrangian (9) is given by
\begin{equation}
{\cal H} (x)  = {\cal H}_\phi(x)+{\cal H}_\chi(x)
+{\cal H}_{I_{1}}(x) + {\cal H}_{I_{2}}(x), 
\end{equation}

where

\begin{equation}
{\cal H}_{\phi}(x)=\dot \phi^{\ast}(x) \dot 
\phi (x) + \nabla \phi^{\ast}(x)  \cdot \nabla \phi (x)
 + m^2 \, \phi^{\ast} (x) \phi (x)          
\end{equation}

\begin{equation}
{\cal H}_{\chi}(x)={1\over{2}}\dot 
\chi_0^2+{1\over{2}}(\nabla \chi_0)^2+
{1\over{2}}\mu^2 \chi_0^2,
\end{equation}

\begin{equation}
{\cal H}_{I_{1}}(x) = g \, \phi^{\ast}(x) \phi (x) \chi_0(x),
\end{equation}

\begin{equation}
{\cal H}_{I_{2}}(x) =-\frac{g^2}{2}\int dx^\prime\, \phi^{\ast}(x)
\phi(x) D(x-x^\prime) \phi^{\ast}(x^\prime) \phi(x^\prime),
\end{equation}
and 

\begin{equation}
D(x-x^\prime)=\int\frac{dk}{(2\pi)^{N+1}}e^{-ik \cdot (x-x^\prime)}
\frac{1}{\mu^2-k \cdot k},
\end{equation}
where $dk = d^{N+1}k$ and $ k \cdot k = k^2= k^{\nu} k_{\nu}$.

To specify our notation, we quote the usual decomposition of the fields in $N+1$ dimension:
\begin{equation}
\phi(x)=\int\,d^N{q}\,\lbrack (2\pi)^N2\omega_q
\rbrack ^{-{1\over{2}}}\lbrack A({\bf q})e^{-i{q \cdot  x}}
+B^{\dagger}({\bf q})e^{i{q \cdot  x}}\rbrack
\end{equation}

\begin{equation}
\chi(x)=\int\,d^N{p}\,\lbrack (2\pi)^N2\Omega_k\rbrack ^{-{1\over{2}}}
\lbrack d({\bf p})e^{-i{p \cdot x}}
+d^\dagger({\bf p})e^{i{p \cdot x}}\rbrack
\end{equation}
where $\omega_p=(p^2+m^2)^{1\over{2}}$, $\Omega_q=(q^2+\mu^2)^{1\over{2}}$,
$ q \cdot x = q^{\nu} x_{\nu}$ and  $q^{\nu} = (q^0=\omega_q, {\bf q})$.  
The momentum-space operators 
$A^\dagger$, $A$, $B^\dagger$, $B$ obey the usual commutation relations. 
The nonvanishing ones are
\begin{equation}
\lbrack A({\bf p}),A^\dagger({\bf q})\rbrack =\lbrack B({\bf p}),
B^\dagger({\bf q})\rbrack =
\delta^N({\bf p}-{\bf q})
\end{equation}
\begin{equation}
\lbrack d({\bf p}),d^\dagger({\bf q})\rbrack 
=\delta^N({\bf p}-{\bf q})
\end{equation}
The Hamiltonian operator, $\ds {\hat H}(t) = { \int d^N x \,{\hat {\cal H}}(x)}$, of the QFTheory
 is expressed in terms of the particle and antiparticle creation and annihilation operators
 $A^\dagger$, $A$, $B^\dagger$, $B$ 
 in the usual way. These operators  are then commuted so that they stand in
  normal order in the Hamiltonian (we are
not interested in vacuum-energy questions in this work). 


\vss \vss
\ni {\bf 3. Ansatz for the particle-antiparticle system, variational equation and effective potentials.}
\vss
The simplest ansatz that can be chosen for a particle-antiparticle ($\phi \phib$) state is
\begin{equation}
\mid \psi_2 \ket=\int\,d^N{ p}_1\,d^N{ p}_2 \, 
F({\bf p}_1, {\bf p}_2)A^\dagger({\bf p}_1)
B^\dagger({\bf p}_2)\mid 0 \ket ,
\end{equation}
where $F$ is an adjustable function.
We use this trial state to evaluate the matrix elements needed to implement the variational 
principle (2), namely
\begin{equation}
\bra \psi_2\mid :{\hat H}_{\phi}-E: \mid \psi_2 \ket  =  
\int d^N{ p}_1 \, d^N{ p}_2 \, 
F^\ast({\bf p}_1,{\bf p}_2)F({\bf p}_1,{\bf p}_2)
\Big[ \omega_{p_1} +\omega_{p_2} -E \Big], 
\end{equation}
and

\begin{eqnarray}
\bra \psi_2 \mid  :  {\hat H}_I: \mid \psi_2 \ket & =&  
\bra\psi_2\mid :{\hat H}_{I_2}: \mid \psi_2 \ket  \nonumber \\
& = & -\frac {g^2}{8(2\pi)^N}\int d^N{ p}_1\, d^N{ p}_2 \, d^N p_1^{\prime}
 \, d^N p_2^{\prime} \, 
F^\ast({\bf p}_1^{\prime} ,{\bf p}_2^{\prime}) F({\bf p}_1,{\bf p}_2) \nonumber\\
& & \times \delta^N({\bf p}_1+{\bf p}_2-{\bf p}^\prime_1-{\bf p}^\prime_2) \,
e^{-i(\omega_{p_1}+\omega_{p_2}-\omega_{p_1^\prime}-\omega_{p_2^\prime}) t} 
\frac{1}{\sqrt{\omega_{p_1}\omega_{p_2}\omega_{p_1^\prime}\omega_{p_2^\prime}}}
\\
& & \times \bigg[ 
\frac{1}{\mu^2-( p_1 +  p_2)^2} +
\frac{1}{\mu^2-( p_1^\prime +  p_2^{\prime})^2 } +
\frac{1}{\mu^2-( p_1^\prime +  p_1)^2 } +
\frac{1}{\mu^2-( p_2^\prime +  p_2)^2} \bigg]  . \nonumber
\end{eqnarray}
 We have normal-ordered the entire Hamiltonian since, at the present level of approximation 
(cf. the trial state (20)),
 this circumvents the need for  mass renormalization which would otherwise arise  
 in eq. (22).  Also,
in the Schr\"odinger picture we can take $t=0$, and  we do so  henceforth.

If we now specialize to the rest frame, where $F(\pb_1, \pb_2) = f(\pb_1) \delta^N(\pb_1 + \pb_2)$, 
then the variational principle (2) leads to the following momentum-space wave equation for the 
relative motion of the particle-antiparticle system:

\begin{eqnarray}
\Big[ 2\omega_{p} -E\Big] f({\bf p})  =  
\frac{g^2 }{4(2\pi)^N}\int d^N{p}^\prime f({\bf p}^\prime)
\frac{1}{\omega_{p}\omega_{p^\prime}} 
  \bigg[ \frac{1}{\mu^2+({\bf p}^\prime-{\bf p})^2 -(\omega_p - \omega_{p^\prime})^2}
 - \frac{1}{4\omega_p^2 - \mu^2} \bigg]
\end{eqnarray}
Note that the kernel (momentum-space potential) in this equation contains two terms. The first 
corresponds to one-chion exchange and the second corresponds to virtual annihilation (this 
is perhaps more obvious from the four manifestly covariant terms  in eq. (22)).

In the nonrelativistic limit, $\ds { {p^2 \over m^2} \ll 1}$, this equation reduces to 
\begin{eqnarray}
\Big[ {\pb^2 \over m} -\epsilon \Big] f({\bf p})  =  
\frac{g^2 }{4(2\pi)^N m^2}\int d^N{p}^\prime\, f({\bf p}^\prime) 
  \bigg[ \frac{1}{\mu^2+({\bf p}^\prime-{\bf p})^2}
- \frac{1}{4m^2-\mu^2} \bigg],
\end{eqnarray}
where $\epsilon = E-2m$. In coordinate space, equation (24) is just the usual time-independent 
Schr\"odinger equation for the relative motion of the particle-antiparticle system:
\begin{equation}
- {{\hbar^2 \over m}}\nabla^2 \psi (\rb) + V(r) \psi (\rb) = \epsilon \psi (\rb) .
\end{equation}
The potential $V(r)$ is a sum of an  attractive Yukawa  potential (due to one-chion exchange)
 and an repulsive (if $\mu < 2m$) contact
potential (due to virtual annihilation). In 3+1 dimensions these are, explicitly, 
\begin{equation}
V(r) = - \alpha {{e^{-\mu r}} \over r} + {{4 \pi \alpha} \over {4 m^2 - \mu^2}}  \delta^3 (\rb),
\end{equation}
where $\ds \alpha = \frac {g^2} {16 \pi m^2}$ is the effective dimensionless coupling constant.

It is clear from Eq. (23) that the relativistic, momentum-space one-quantum exchange potential is 
always attractive, while the virtual annihilation potential is repulsive if the mass   $\mu$ of the 
mediating-field quantum is not too large, namely if $\mu \le 2 m$. However, if $\mu > 2 m$ then 
the annihilation potential becomes attractive at low momenta.


\vss 

\ni {\bf 4. Analytic solution of the variational particle-antiparticle equation with virtual annihilation  
interaction only.}
\vss
The variational particle-antiparticle equation (23) cannot be solved analytically. Of course, numerical 
solutions can be obtained, and these will be discussed in section 6. In addition, for the massless-exchange 
case, analytic perturbative solutions can be worked out, and these will be presented in section 5. 
However, if only the annihilation interaction is kept (i.e. the chion-exchange 
interaction is turned off), exact analytic solutions of such a simplified particle-antiparticle equation can
 be obtained (for both bound and scattering states). This is of interest as a solvable relativistic 
two-body equation, if
for no other reason.  Thus, if we neglect the first interaction term (the one-quantum-exchange term)
 in eq. (23) we obtain the rest-frame ``annihilation-interaction'' equation

\begin{equation}
[2 \omega_p - E] f(\pb) = f_0(\pb)+{{g^2} \over {4 (2 \pi)^N}} \int d^N\pp f(\pbp) 
{1 \over {\omega_p 
\omega_{\pp}}} {1 \over {\mu^2 - 4 \omega_p^2}},
\end{equation}
where $f_0(\pb) $ is a ``plane wave'' solution of eq. (27) ( with $g=0$) representing the 
particle and antiparticle 
incident on each other with energy $E = 2 \omega_{p_0}$. Of course, $f_0(\pb) =0$ for bound states.
Because the annihilation interaction is entirely repulsive  if $\mu \le 2 m$, we have only 
particle-antiparticle scattering solutions  in that case.

We shall discuss the solution of eq. (27) in some detail only in 3+1 dimensions.  It is evident that the 
integral on the right-hand-side of eq. (27) vanishes, except in $S$-states, hence is sufficient to 
write down the $S$-wave component of eq. (27), namely
\begin{equation}
f(p) = {1 \over {4 \pi p^2}} \left[ \delta (p-p_0) - { {\alpha \,{m^2} A\, p^2} \over { (\omega_p - \omega_{p_0}) 
( \omega_p^2 - ({\mu \over 2})^2 )\, \omega_p} } \right] ,
\end{equation}
where 
\begin{equation}
A = \int_0^{\infty} dp\, p^2 {1 \over \omega_p} f(p) .
\end{equation}
From equation (28) it follows that the $S$-wave phase shift $\eta$ is given by
\begin{equation}
\tan \eta =  - \pi \alpha \, m^2 A {p_0 \over { \omega_{p_0}^2 - (\mu / 2)^2 }} .
\end{equation}
Substitution of eq. (28) into eq. (29) yields the result
\begin{equation}
A= {1 \over {\omega_{p_0} \left( 4 \pi + \alpha \, I(m,\mu,p_0) \right)} } ,
\end {equation}
where 
\begin{eqnarray}
I(m,\mu,p_0)  = m^2 \,  {\cal P} \int_0^\infty dp {p^2 \over {( \omega_p - \omega_{p_0}) \, 
\omega_p^2  \, \left(\omega_p^2 - (\mu/2)^2\right) }} .
\end{eqnarray}
This principal value integral is, explicitly,
\begin{eqnarray}
I(b,q) = & &{1 \over {2(q^2+b^2)}} \Bigg\{{b \over \sqrt{1-b^2}} \ \left[ \pi - 2 \tan^{-1} 
\left( {b \over \sqrt{1-b^2}} \right) \right] \nonumber \\
& & -\, {q \over \sqrt{1+q^2}} \left[ 2 \tanh^{-1} 
\left( {q \over \sqrt{1+q^2}}\right) + { \pi \, q \over {b+1}} - { \pi \, b \over{q(b+1)}} \right] \Bigg\}
\end{eqnarray}
if  $0<b<1$, where $b^2 = 1- (\mu/2m)^2 $, $q=p_0/m$, and
\begin{equation}
I(q) = {1 \over {q^2+1}} +  {1 \over {(q^2+1)^{3/2}}} \Big[ {\pi \over 4} (1-q^2) 
-q \, \tanh^{-1} 
\left( {q \over \sqrt{1+q^2}}\right)  \Big]
\end{equation}
for the massless exchange case, $\mu =0$ (i.e. $b=1$).
With $A$ as given in eq. (34), the tangent of the $S$-wave phase shift becomes
\begin{equation}
\tan \eta =  - \, { {\pi \alpha \, m^2 \, p_0} \over {\omega_{p_0} \left( \omega_{p_0}^2 - (\mu / 2)^2 \right) 
\big( 4 \pi + \alpha \, I(m,\mu,p_0) \big)} } ,
\end{equation}
from which the elastic particle-antiparticle scattering cross section $\ds \sigma = { {4 \pi}
 \over {p_0}^2} \sin^2 \eta $ is readily calculated.

For the massless chion exchange case ($\mu = 0$), the cross section, in units of $\pi/m^2$, starts from
 a value of $\ds {{64 \alpha^2} \over {(16 +[1+4/\pi] \alpha)^2}}$ at $q=p_0/m=0$, then decreases 
 monotonically  with increasing $q$ to the asymptotic form  $ \ds { \alpha^2 \over 
{ 4 q^6}} \to 0$ as $q \to \infty$. 
 Note that the maximum value of the cross section (which occurs at $q=0$ for all $\alpha$
 in this massless-exchange case) 
increases uniformly from zero at $\alpha=0$ to an asymptotic value of $64 \pi^2 /(\pi +4)^2 = 12.3848$ 
as $\alpha \to \infty$. 

When the mediating field quanta are massive then, for given $\alpha$, the cross-section, 
as a function of collision energy, behaves qualitatively in a similar way to the massless case if $\mu/m < 1$.
 However the shape of the cross section changes as $\mu$ increases towards $2m$, in that it
 initially increases with the collision energy, reaches a maximum and then decreases towards zero as
\begin{equation}
{ \sigma \over {\pi /m^2}} \sim  {1 \over 4} {\alpha^2 \over q^6} + {1 \over 16} {\alpha^3 \over { (b+1) q^7}}
 + O({1 \over q^8}) .
\end{equation} 
We shall not delve into a detailed discussion of the behaviour of $\sigma (q)$ for various $ \alpha$ 
and $ b$ since the 
specific results can always be evaluated using the given analytic formulae (33 - 35).

It is of interest to note that in the non-relativistic limit, i.e. $p \ll m$,  eq. (27) becomes, 
\begin{equation}
\left [{p^2 \over m} - \epsilon \right] f(\pb) = f_0(\pb) +{g^2 \over {4 (2 \pi)^N}}
{1 \over {m^2(\mu^2 - 4 m^2)}} \int d^N\pp f(\pbp)  ,
\end{equation}
where $\epsilon = E -2 m$.
The resulting  $I$ integral (cf. eq. (32)) diverges in $N=2$  and 3 dimensions,
 thus resulting in a vanishing phase shift and 
cross section.  This just reflects the ``trivial'' nature of the  scattering by a repulsive delta function potential 
in non-relativistic (Schr\"odinger) theory in $N > 1$ dimensions
(``trivial'' in the sense  that the $S$-matrix is unity).

For very massive mediating fields, $\mu > 2m$, the annihilation interaction  becomes attractive at 
low momenta, and this leads to binding of the particle-antiparticle system if the coupling constant 
is large enough. The energy eigenvalue condition is (from eq. (27) with $f_0 = 0$) 
\begin{equation}
1=  {{g^2} \over {4 (2 \pi)^N}} \int d^Np\,
{1 \over {(2 \omega_p -E) \,
\omega_p^2  \, (\mu^2 - 4 \omega_p^2)}} .
\end{equation}
In $N=3$ dimensions equation (38) yields the result
\begin{eqnarray}
{{2 \pi} \over \alpha} &=& { 2 \over {4+4 b^2- \ep^2} }  
 \Bigg\{   { b \over \sqrt{1+b^2} }  
\tanh^{-1} \left(  { b \over \sqrt{1+b^2} }  \right)   +   { \sqrt{4-\ep^2} \over \ep }
 \tan^{-1} \left(  { \sqrt {4-\ep^2} \over \ep }  \right)  \nonumber  \\
& & +\, \pi { b^2 \over {1+b^2} }    { \ep \over \sqrt{4 - \ep^2} }  \Bigg\}  
+ {\pi \over 2} { 1 \over {1+b^2}} {1 \over \ep} \left( 1 - {4 \over  \sqrt{4 - \ep^2}}
\right) ,
\end{eqnarray}
where, now, $b^2 = (\mu/2m)^2-1$ and $\ep = E/m$.  A plot of  $E(\alpha)$ for a representative value 
of $b=1$, i.e. $\mu = 2 \sqrt{2}\, m$, is given in Figure 1 (actually, $ \alpha (E)$ is plotted). Note that 
binding does not set in until the coupling constant $\alpha$ exceeds a minimum value $\alpha_0$. 
Thereafter the energy (i.e. particle-antiparticle mass) decreases monotonically with increasing 
$\alpha$ to an asymptotic value $E/m = \ep_{ \rm min}$, where both $\alpha_0$ 
and $\ep_{\rm min}$ vary with $\mu/m$.  The general expression for $\alpha_0$
is
\begin{equation}
\alpha_0 = 2 \pi \Bigg[ {1 \over 2} {1 \over {b \sqrt{b^2+1}}} \tanh^{-1}\left( {b \over \sqrt{b^2+1}} 
 \right) + {\pi \over 4} { 1 \over {b^2+1}} \Bigg]^{-1}
 \sim  {{8\pi} \over {\pi + 2}} + {{8\pi(4+3\pi)} \over {3(\pi+2)^2} } b^2 + O(b^4) .
\end{equation}
 The minimum value of $\alpha$ at which binding sets in (vis. $\alpha_0 = {{8\pi} \over {\pi + 2}} =
4.88812$)  occurs when $b=0$ (i.e. when $\mu$ just passes $2m$), but the binding is very weak, since  
$\ep_{\rm min}$ is barely below 2 in that case.  As $b$ increases, so does $\alpha_0$ and so does
 the binding energy.  For $\alpha$ near $\alpha_0$ the behaviour of $E(\alpha)$ is of the form
\begin{equation}
{E \over m} = 2 - \left( {{4b^2} \over \alpha_0} \right)^2 (\alpha - \alpha_0)^2 + O((\alpha-\alpha_0)^3) .
\end{equation}

Once again, it is of interest to note that the annihilation interaction,  in $N=2,3$ spatial dimensions, 
can support bound states only in the relativistic formulation. 
 This is because, had we started from the non-relativistic form of eq. (37) (with $f_0=0$
 for bound states),  then the eigenvalue equation corresponding to equation (38) would be
\begin{equation}
1=  {{g^2} \over {4 (2 \pi)^N}} {1 \over { m^2 (\mu^2 - 4 m^2)}} \int d^Np\,
{1 \over {(p^2/m -\epsilon) }} ,
\end{equation}
The integral (42)  converges only for $N=1$, whereas in the relativistic case 
the  integral (38) converges for $N=1,2,3$.  In other words the short-range annihilation 
interaction does not support bound states non-relativistically in 3 or even 2 spatial dimensions  
(the interaction is an attractive delta-function potential in the non-relativistic limit), but it 
does support bound states  relativistically. This implies that the virtual annihilation interaction
 strengthens 
(relative to the non-relativistic delta function potential) if relativity is taken into account.


\vss \vss
\ni {\bf 5. Perturbative results in 3+1 dimensions for the particle-antiparticle binding energy in the 
massless-exchange case.}
\vss

The relativistic two-particle equation (23) can be reduced to radial form 
by setting
\begin{equation}
f({\bf p})=f(p)Y_{\ell m}(\hat{\bf p}),
\end{equation}
where $p=|{\bf p}|$ and $Y_{\ell m}(\hat{\bf p})$ are the usual spherical harmonics,
and carrying out the angular integration. The result,   
 in $N=3$  dimensions, is
\begin{equation}
\Big[ 2\omega_{p} -E\Big] f(p) =\frac{\alpha}{\pi}\int_0^\infty
dp^\prime\frac{p^\prime}{p}f(p^\prime)K_\ell (p^\prime, p) ,
\end{equation}
with 
\begin{equation}
K_\ell (p^\prime, p)=\frac{m^2}{\omega_p\omega_{p^\prime}} \bigg[
Q_\ell (z)-\frac{\pi}{ \omega_p^2-({\mu \over 2})^2}p\,p^\prime\, \delta_{\ell 0} \bigg] ,
\end{equation}
where
\begin{equation}
z=\frac{p^2+p^{\prime2}+\mu^2-(\omega_p-\omega_{p^\prime})^2}{2pp^\prime} ,
\end{equation} 
${\ds \alpha = \frac {g^2} {16 \pi m^2}}$, and $Q_\ell(z)$ is the Legendre function of the second kind.
This equation is similar to that derived by Di Leo and Darewych using a variational-perturbative 
approach [33].  That previous result corresponds to eq. (45) without the virtual-annihilation 
interaction (the second term in eq. (45)), and with $\omega_p=\omega_{p^\prime}$ in the first,
one-chion exchange term of eq. (45).

Since the solutions of equation (44) in the non-relativistic limit (without the virtual annihilation
 interaction) are the well
known hydrogenic wave-functions in momentum-space [34], we can use them
to obtain perturbative expressions to the particle-antiparticle mass (rest-energy). The result is

\begin{eqnarray}
E_{n \ell}(\alpha) & =  & 2m -\frac{1}{4} m \alpha^2 \frac{1}{n^2}
-\frac{1}{16}m\alpha^4 \bigg[ \frac{2}{(2 \ell+1)n^3}-\frac{3}{4n^4} \bigg]
\nonumber \\
& & \qquad + \frac{1}{8}m\alpha^4\bigg[ \frac{4}{(2 \ell+1)n^3}-\frac{1}{n^4} \bigg]
+\Delta E_{anni} +O(\alpha^5) .
\end{eqnarray}
The terms on the right are the rest energy, the non-relativistic Balmer term, the O($\alpha^4$) 
correction to the kinetic energy, one-chion exchange interaction and the virtual annihilation 
interaction, respectively.
The correction due to the annihilation interaction is
\begin{eqnarray}
\Delta E_{anni} & = & \frac{\alpha}{\pi} m^2 \int_0^\infty\int_0^\infty
dp\,dp^\prime f(p)f(p^\prime)\frac{1}{2\omega_p^3\omega_{p^\prime}}p^2
\,p^{\prime 2}\, \delta_{\ell 0}
\nonumber \\
& =  &  \frac {1}{8}  m \alpha^4 \frac{1}{n^3}\delta_{\ell 0} +O(\alpha^5)
\end{eqnarray}
The result  (47) agrees with the earlier work [33], except that  the annihilation correction (48) is new.
The virtual annihilation correction is reminiscent of what is obtained for triplet $\ell =0$ states of 
positronium, where one obtains $ \ds {1 \over 4} m \alpha^4 {1 \over n^3} \delta_{\ell 0} 
\delta_{S 1}$, with $\alpha = e^2/\hbar c$ in that case. Note that the ``retardation term'', 
$(\omega_p - \omega_{p^\prime})^2$ of Eqs. (23) and (46), has no effect at $O(\le \alpha^4)$.

 As is well known, the massless Wick-Cutkosky model has 
been solved 
in the ladder approximation of the Bethe-Salpeter formulation [17, 18], as well as the light-cone 
ladder approximation [19, 20]. The expansion  of these solutions 
 in powers of $\alpha$  is found to be [19]
\begin{equation}
E /m = 2 - {\alpha^2 \over {4 n^2}} - {{\alpha^3 \ln \alpha} \over {\pi n^2}} + O(\alpha^3)
\end{equation}
This is quite different from our result (47) beyond the $O(\alpha^2)$ Balmer term. The unusual
$O(\alpha^3 \ln \alpha)$ terms are an artefact of the {\sl ladder} Bethe-Salpeter 
formulation, and allegedly do not arise if crossed-ladder diagrams are included  [35]. A  
discussion of the origin of the $\alpha^3 \ln \alpha$ term is given by A. Amghar and B. Desplanques [36].
In any case, our results 
contain no such terms, and are much more like the corresponding results for positronium in this 
respect (i.e. that the lowest order relativistic corrections to the Balmer result are $O(\alpha^4))$.

\vss \vss
\ni{\bf 6. Numerical solution for the $\mu / m = 0$ and $\mu / m = 0.15$ cases.}
\vskip .3cm
We have solved equation (44) approximately for the ground state in $N=3$ spatial dimensions, 
using the variational method with the trial wave function 
$\ds f(p) = \frac { \omega_p} {(p^2+b^2)^n}$, where  $b$ is an  adjustable parameter, determined 
by minimizing 
\begin{equation}
  E =  \left[ { \int_0^\infty dp \, 2\omega_{p} \,  p^2  f^2(p) -\frac{\alpha}{\pi}\int_0^\infty \int_0^\infty
dp^\prime dp \,{p^\prime}{p}f(p^\prime) f(p) K_\ell (p^\prime, p)}  \right]
\Bigg /  { \int_0^\infty dp \,  p^2  f^2(p)}    ,
\end{equation}
with respect to $b$, for various $n$ and given $\alpha$.  Although these variational results are only 
approximations to the ``exact'' (i.e. numerical) solutions of eq. (44), they are in fact 
 reasonably close to the numerical ones for the entire range of values of $\alpha$ considered, 
as shown in a previous study [33]. A list of ground-state
 values of $E/m$ in the massive-exchange case for the present model is given in Table 1, for $n=2, 2.5, 3,
 3.5$. Generally, the $n=2$ values are lowest for $\alpha < 0.5$, where relativistic effects are not
 so pronounced, but the $n=2.5$ and $n=3$ values are lower at strong coupling. However, 
the various values look quite similar on a graph, and in the figures we shall plot 
curves corresponding to a single value of $n$ only, as this will be  
 sufficient for comparison purposes.  

An advantage of using the variational solution is that we 
have an analytic representation of the wave-function, and so can examine its behaviour as 
the coupling constant $\alpha$ changes. Thus, we note that the values of the ``inverse Bohr radius''
 parameter $b$, for any given $n$, increase monotonically with increasing $\alpha$.  Also, for any
given value of $\alpha$, the parameter $b$ increases as $n$ increases.

A plot of  $E(\alpha, \mu=0)$ for the ground state obtained
 in this way is 
shown in Figure 2. We plot two versions of our results, namely with and without the virtual annihilation 
interaction (second term of eq. (45)) included. Note that the affect of virtual annihilation is 
substantial, and it increases with increasing $\alpha$.  

As mentioned previously,
 the scalar Yukawa (or Wick-Cutkosky) model has been studied by many authors in various
 formalisms and approximations.  It is therefore of interest to compare 
some of them to our results. Thus Figure 2 also contains plots of the classic solutions of 
Wick [17] and Cutkosky [18] of the Bethe-Salpeter equation in the ladder approximation, as well 
as the analogous light-cone calculations of Ji and Furnstahl [20]. We also plot the results of 
 Di Leo and Darewych [33], which correspond to the present results  with 
$\omega_p = \omega_{p^\prime}$
 (i.e. no retardation).  None of these results contain the virtual
 annihilation interaction (which is 
repulsive) and so they should be compared with the present results without virtual annihilation.

It is evident from Fig. 2 that our variational results predict stronger binding that either of the 
ladder Bethe-salpeter results, or the Di Leo results. (The difference between our no-virtual-annihilation
 results and the Di Leo results show the effect of the retardation term
 $(\omega_p - \omega_{p^\prime})^2$ in the potential of eq. (23).) Indeed, our results predict 
stronger binding 
 than  ladder B-S, even if we include the repulsive virtual annihilation interaction.  Numerical 
values corresponding to Figure 2 are listed in Table 2 for $n=2$, along with the wave-function
 parameter $b$.   
Table 3 is a list of corresponding results for $n=3$.

It has been shown recently [37] that exact two-body eigenstates can be written down 
 for the QFTheoretic Hamiltonian 
$H_\phi +H_{I_2}$ (cf. eqs. (10, 14)), that is, for the present model without free chions, provided 
that an ``empty'' vacuum state $| \tilde 0 \ket$, annihilated by both positive and negative 
 energy components of the field operator $\phi (x)$, is used.  The use of such an 
empty vacuum state results in a relativistic 
 two-body scalar equation that has both positive and negative energy solutions.  Such negative-energy 
solutions do not arise (and should not arise) in the conventional QFT treatment that uses a Dirac 
``filled-negative-energy-sea'' vacuum state, including the present work. Nevertheless the two-body 
 equation obtained in [37] can be solved analytically [37, 38] in the massless-mediating-field case, 
and the positive energy, `$E \simeq 2m$' - like solution is
\begin{equation}
E = m \sqrt{2 \left (1 + \sqrt {1-\aln^2}\right )} = m \left(2 - 
{1\over 4} {\aln}^2 - {5\over{64}} \aln^4 + O(\alpha^6)  \right)\; . 
\end{equation}
This result is quite different from that of the present treatment, or the ladder Bethe-Salpeter 
calculations. For one thing there is a rather low critical value of the coupling constant, namely $\alpha_c = n$ 
($\alpha_c = 1$ for the ground state) beyond which the two-body energy (rest mass) ceases to 
be real ($E/m = \sqrt {2} = 1.414...$  when $\alpha = \alpha_c$). This is similar to what happens 
for one-body Klein-Gordon or Dirac equations in a Coulomb potential.  We also plot the result (51) 
in Fig. 2, and note that in the domain $0 \le \alpha \le 1$ it predicts stronger binding than any of 
the other results shown in the figure. 

(An aside: The formula (51), and its $m_1 \ne m_2$ generalization derived in ref. [38], was
 obtained previously by Todorov [25] using a quasipotential approach. Todorov's article [25],
 which came to our attention recently, also contains a useful historical overview of earlier work on the 
relativistic two-body problem in QFT, including references to various ``rediscovered'' formulations 
and results.)

Figure 3 is a plot of the particle-antiparticle ground state energy in $N=3$ spatial dimensions for the massive 
exchange case, with $\mu / m = 0.15$. Once again, we plot the solutions of our variationally derived 
equation (44) with and without virtual annihilation included. In addition, we plot some Bethe-Salpeter 
based quasipotential results that are given in a study of the $\phi^2 \chi$ model by Nieuwenhuis and Tjon [27b], 
together with their numerical results obtained using a Feynman-Schwinger formulation.  The various 
quasipotential results plotted in Figure 3 are explained in the paper of Nieuwenhuis and Tjon [27b], 
and this will not be repeated here. None of the calculations, save ours, contain the virtual 
annihilation interaction, so they should be compared to the version of our results without virtual annihilation.
Some numerical values corresponding to the curves plotted in Figure 3 are listed in Table 4.

It is evident from Figure 3 that the ladder Bethe-Salpeter (BS) calculation predicts the weakest binding, 
while the Feynman-Schwinger (FS) results of Nieuwenhuis and Tjon predict the strongest binding. 
(The FS calculations are nonperturbative, but contain no loop effects.)
The various quasipotential results are distributed between the ladder BS and the FS values, in the 
following order: 
Blankenbecler-Sugar-Logunov-Tavkhelidze [39],  Gross 
(with retardation) [40] and ``equal-time'' [41].
All these quasipotential results are taken from Figure 1 of Nieuwenhuis and Tjon [27b]. Our present
results (without virtual annihilation) fall above most of the quasipotential ones, and lie closest to 
the Blankenbecler-Sugar curve.  We also include a plot of the two-body Klein-Gordon Feshbach-Villars 
formalism results of ref. [37], which lie very close to the Gross results.
 We can only speculate to what extent this is a coincidence (the Gross and FV formalism equations are 
not obviously similar). 

Theussl and Desplanques [42] have recently calculated the two-body energy being studied here. They use 
the Bethe-Salpeter equation, but include crossed-ladder effects in an approximate perturbative way. Their 
results fall very close to the Gross quasipotential values (and to those of the Feshbach-Villars formulation)
 in the domain $\alpha < 1$. Therefore, we do not plot them in figure 3.  The Theussl-Desplanques work 
underscores the inaccuracy of the ladder Bethe-Salpeter approximation and the importance of including 
crossed-ladder effects. 

The fact that the present variational results show considerably weaker binding than the FS results 
of Nieuwenhuis and Tjon is perhaps not surprising, since the present calculation contains no
``crossed-ladder'' effects (the  simple variational ansatz (20) is incapable of incorporating such 
effects).  However, our results show considerably stronger binding than the ladder BS, even though 
the latter uses essentially the same kernel (one-chion exchange) as the present calculation.

The quasipotential results differ from our variationally derived values in several respects.  For one
thing, the quasipotential equations are all different and somewhat ad-hoc (though physically
 motivated) one-time modifications of the BS equation, whereas our equations are 
obtained in a  completely ``ab-initio'' way, and
 are limited primarily by the  simple choice of ansatz (20) that we have made 
in this work. The quasipotential equations, in the 
unequal mass case, have the Klein-Gordon (KG) equation as their one-body limit. In this sense they are 
more like the two-body KG Feshbach-Villars  results of refs. [37, 38], which also have 
the KG eq. as their one-body limit.  The present variational approach leads to an equation, which, in 
the unequal mass case, does {\sl not} have the KG one-body limit.  

The stipulation that the ``correct'' 
one-body limit of a relativistic two-scalar-particle equation should be the KG equation is often made in the 
literature.  However, this is a curious measure of correctness, since the KG equation has 
negative-energy solutions, which should not arise in a conventional QFT treatment that uses a Dirac
``filled-negative-energy sea'' vacuum.  Indeed, negative energy solutions cannot, and should not, 
arise in the present calculations (we use the conventional Dirac vacuum), and so the unequal-mass 
counterpart of eq. (27) does not (and should not) have the KG equation as its one-body limit.  On the other 
hand, the two-body KG Feshbach-Villars formalism equation (given in ref. [38] for the unequal mass 
case) is obtained using an ``empty'' vacuum, and so it does have negative-energy solutions, 
and also the KG one-body limit. Its 
binding energy predictions are similar to those of the  quasipotential equations, even though it contains 
no retardation, or manifest crossed-ladder effects.  The one-body limit equation in the present formalism 
has only positive-energy solutions, like the two-body equation (27). 

Of course, we do not claim that our present results are ``better'' than the quasipotential equation 
ones (in the sense that they are closer to the unknown exact results for this model).  The trial state 
(20) that we are using here is  too simple to make any such claim. However, our approach is
strictly variational, with nothing ``put in by hand'', and can be systematically improved by improving 
 on the trial state (20) (which we are in the process of doing).  It may be that the numerical FS results of 
Nieuwenhuis and Tjon are the most accurate binding-energy results available to date for the 
scalar Yukawa model,
so it will be interesting to see how improved variational results will compare with this benchmark.

We should point out that the trial state (20) that has been used in this work is insensitive to the $H_{I_1}$ 
term of the Hamiltonian (cf. Eq. (13)). Thus, it is suitable for describing stable particle-antiparticle 
states only, without explicit annihilation of the particle-antiparticle system, or decay of the excited 
bound states, with the emission of {\it physical} chions.  It is possible to include such processes 
in the present formalism in various ways, such as perturbatively, or by suitable modification 
of the ansatz (20) (see, for example, refs. [33], [43]). 
However,  we do not consider such processes in this paper. 
 \vfill  \eject


\ni {\bf 7. Concluding remarks.}
\vss
We have applied the variational method to the study of particle-antiparticle bound states in the
 scalar Yukawa  
(Wick-Cutkosky) model (scalar particles interacting via a massive or massless mediating 
 scalar ``chion'' 
field).  We have used a reformulated version of this theory in which a covariant  Green function 
is used to eliminate the chion field partially, so that the chion propagator appears directly in the 
QFTheoretic hamiltonian.

A simple Fock-space trial function is used in the variational method. It leads to a relativistic 
particle-antiparticle momentum-space equation with the covariant one-chion exchange and 
virtual annihilation Feynman amplitudes appearing in the kernel (momentum-space potential) 
of the equation.  The virtual annihilation interaction is repulsive, except if the mediating-field quantum 
(the chion) is very heavy (more massive than the combined rest mass of the particle and 
antiparticle) whereupon it becomes attractive at sufficiently large momenta.  This particle-antiparticle 
equation has no negative energy solutions, i.e. it is free of any negative-energy ``pathologies''.

When the one-chion exchange interaction is turned off, we find that the resulting model 
theory (with a purely virtual-annihilation interaction) is analytically solvable for 
scattering states, and also for bound states when the chions are very massive. The virtual 
annihilation interaction reduces to a delta-function (contact) potential in the nonrelativistic
 limit. It supports no bound states in $N=3$ spatial dimensions and  the $S$-matrix 
is unity in this limit.  However, if the relativistic equations are used, we find that bound states are possible 
and the scattering is not trivial. This analytically solvable relativistic model is instructive in 
understanding the effects of a relativistic generalization of the delta-function potential, such as occurs 
in the virtual annihilation interaction.

In the general case, with both one-chion exchange and virtual annihilation interactions 
included, the relativistic particle-antiparticle equation cannot be solved analytically (at least, we
do not know how to do so). However, analytic expressions for the energy (rest mass) of the 
bound particle-antiparticle system can be obtained as an expansion in the effective 
dimensionless coupling constant $\alpha$ for the case of a massless mediating field. 
 We give such an expression for the ground and arbitrary excited states of the system to 
O($\alpha^4$) inclusive.  We find that the lowest-order relativistic corrections to the 
Balmer formula are O($\alpha^4$), much like for positronium, and quite unlike the predictions 
of the Bethe-Salpeter equation in the ladder approximation, which include unusual
 $\alpha^3 \ln \alpha$ terms.

Lastly, we calculated the particle-antiparticle bound-state energy for arbitrary $\alpha$ in the 
ground state. This was done using a variational approximation rather than numerical integration 
of the two-body equation, as the results are not much different and the variational method allows 
one to exhibit the behaviour of the wave-function more transparently for various values of $\alpha$.
We study the case where the exchanged quantum has mass $\mu = 0$, and find that our results 
predict stronger binding than ladder Bethe-Salpeter approximations (Wick-Cutkosky or light-front 
solutions).  For the case of massive chion exchange, with $\mu = 0.15$, we find analogous behaviour.
However, our results show weaker binding than most of the quasipotential reductions of the 
Bethe-Salpeter equations.

The present approach has several attractive features.  Firstly, it leads to equations 
with no negative-energy or 
 mixed-energy solutions, such as arise in many other formulations (though such negative-energy 
and mixed-energy solutions usually are not  discussed). Secondly, the results are strictly variational, 
with no perturbative approximations. Thus the results are applicable at 
strong coupling, at least in principle (we hasten to add that variational results are only 
as good as the trial states 
employed in their use).  Unlike the quasipotential reductions of the Bethe-Salpeter equation, our results 
are rigorous in the sense that nothing is put in by hand.  Thirdly, the method is amenable to systematic
 improvement by improving the variational trial state.  
Lastly, the method is straightforwardly generalizable 
 to relativistic three or more particle systems, though, of course, one is then faced with the usual 
complexity of a relativistic many-body problem [44].

We thank B. Desplanques for sending us preprints of recent work, and for useful conversation. 
The support of the Natural Sciences and Engineering Research Council of Canada for this work is 
gratefully acknowledged.

\vss 

\ni {\bf References}
\medskip
\begin{enumerate}
\item L. I. Schiff, Phys. Rev. {\bf 130}, 458 (1963).
\par
\item S. Coleman and E. Weinberg, Phys. Rev. D {\bf 7}, 1888 (1973)
\par
\item S. Coleman, Phys. Rev. D {\bf 11}, 2088 (1975).           
\par
\item R. Jackiw, Phys. Rev. D {\bf 9}, 1686 (1974)
\par
\item T. Barnes and G. I. Ghandour, Phys. Rev. D {\bf 22}, 924 (1980).          
\par
\item P. M. Stevenson, Phys. Rev. D {\bf 30}, 1712 (1984) and {\bf 32}, 1389 (1985), 
and P. M. Stevenson and I. Roditi, {\bf 33}, 2305 (1986).               
\par
 \item R. Tarrach, Phys. Letters {\bf B262}, 294 (1991); R. Munoz-Tapia 
and R. Tarrach, Phys. Letters {\bf B256}, 50 (1991).
\par
\item  P. Cea, Phys. Lett. {\bf B165}, 197 (1985);  M. Consoli and G.
 Preparata, Phys. Lett. {\bf B154}, 411 (1985).  See also 
 {\sl Proceedings of the International Workshop on Variational 
Calculations in Quantum Field Theory}, Wangerooge, Germany, 
1-4 September 1987. L. Polley and D. E. L. Pottinger, eds. World Scientific, Singapore, 1988.
\par
\item J. W. Darewych, Ukr. J. Phys. {\bf 41}, 41 (1996).
\par
\item U. Ritschel, Z.  Physik C {\bf 47}, 475 (1990); R. Ibanez-Meier {\sl et al.}
 Phys. Rev. D {\bf 45}, 2893 (1992).
\par
\item I. Tamm, J. Phys. USSR {\bf 9}, 449 (1945).
\par
\item S. M. Dancoff, Phys. Rev. {\bf 78}, 382 (1950).
\par
\item W. Thirring, Ann. Phys. (N.Y.) {\bf 3}, 91 (1958); see, also, ref. [3].
\par
\item J. Schwinger, Phys. Rev. {\bf 128}, 2425 (1962).
\par
\item J. W. Darewych, M. Horbatsch and R. Koniuk, Phys. Rev. Lett. {\bf 54},
 2188 (1985); J. W. Darewych, M. Horbatsch and R. Koniuk, in
{\sl Proceedings of the International Workshop on Variational 
Calculations in Quantum Field Theory}, Wangerooge, Germany, 
1-4 September 1987. L. Polley and D. E. L. Pottinger, eds. World Scientific, Singapore, 
1988, p. 172; Phys. Rev. D {\bf 39}, 499 (1989).
\par
\item  J. W. Darewych, Annales de la Fondation Louis de Broglie {\bf 23}, 15 (1998); and 
J. W. Darewych, in {\sl Causality and Locality in Modern Physics}, G. Hunter {\sl et al.} (eds.), 
Kluwer, Dordrecht, 1998, pp. 333-244.
 \par

\item G. C. Wick, Phys. Rev. {\bf 96}, 1124 (1954).
\par\noindent
\item R. E. Cutkosky, Phys. Rev. {\bf 96}, 1135 (1954).
\par\noindent
\item G. Feldman, T. Fulton and J. Townsend, Phys. Rev. D {\bf 7}, 
1814 (1973).
\par\noindent
\item C.-R. Ji and R. J. Funstahl, Phys. Lett. {\bf 167B}, 11 (1986);  C.-R. Ji, Phys. Lett. B 
{\bf 322}, 389 (1994)
\par\noindent
\item L. M\"uller, N. Cimento {\bf 75}, 39 (1983)
\par\noindent
\item S. J. Brodsky, C.-R. Ji and M. Sawicki, Phys. Rev. D {\bf 32}, 1530 (1985)
\par\noindent
\item C. Schwartz, Phys. Rev.  {\bf 137}, B717 (1965); C. Schwartz and C. Zemach {\bf 141}, 
1454 (1966).
\par\noindent
\item G. B. Mainland  and J. R. Spence,  Few-Body Systems 
{\bf 19}, 109 (1995).
\par\noindent
\item I. T. Todorov, in {\sl Properties of the Fundamental Interactions}, A. Zichichi, ed. 
 Editrice Compositori, Bologna, 1973 Part C, pp. 953-979.
\par\noindent
\item J. Bijtebier and J. Broekaert, J. Phys. G {\bf 22}, 1727 (1996)
\par\noindent
\item T. Nieuwenhuis and J. A. Tjon, Few-Body Syst. {\bf 21}, 
167 (1996); Phys. Rev. Lett. {\bf 77}, 814 (1996).
\par\noindent
\item B. Hamme and W. Gl\"ockle, Few Body Systems {\bf 13}, 1 (1992)
\par\noindent
\item V. A. Rizov, H. Sazdjian and I. T. Todorov, Ann. Phys. {\bf 165}, 59 (1985)
\par\noindent
\item R. Rosenfelder and A. W. Schreiber, ``Polaron Variational Methods in Relativistic Field Theory", 
in {\sl Proceedings of the Dubna Joint Meeting: Int. Seminar on Path Integrals and Applications 
and 5th Int. Conf. on Path Integrals from MeV to GeV}, V. S. Yarunin and M. A. Smodyrev, eds. 
JINR, Dubna, Russia (1996).
\par\noindent
\item N. Nakanishi, Prog. Theor. Phys. Suppl. No. 95  (1988)
\par\noindent
\item Z. K. Silagadze, Budker Inst. of Nucl. Phys. preprint, hep-ph 9803307v2 (1998)
\par\noindent
\item  L. Di Leo  and J. W. Darewych,  Can. J. Phys. {\bf 70}, 412 (1992).
\par\noindent
\item  H. A. Bethe and E. E. Salpeter, {\sl Quantum Mechanics of One- and Two-Electron Atoms}, 
Springer-Verlag, Berlin 1957.
\par\noindent
\item J. L. Friar, Phys. Rev. C  {\bf 22}, 796 (1980).
\par\ni
\item A. Amghar and B. Desplanques, Few Body Systems, in press (1999)
\par\ni
\item J. W. Darewych, Can. J. Phys.  {\bf 76}, 523 (1998).
\par\ni
\item M. Barham and J. Darewych, J. Phys. A  {\bf 31}, 3481 (1998).
\par\ni
\item  A. A. Logunov and A. N. 
Tavkhelidze, N. Cimento {\bf 29}, 380 (1963); 
R. Blankenbecler and R. Sugar, Phys. Rev.  {\bf 142}, 1051 (1966).
\par\ni
\item F. Gross, Phys. Rev. C  {\bf 26}, 2203 (1982).
\par\ni
\item S. J. Wallace and V. B. Mandelzweig, Nucl. Phys.   {\bf A503}, 673 (1989); 
E. Hummel and J. A. Tjon, Phys. Rev. C {\bf 49}, 21 (1994); 
P. C. Tiemeijer and J. A. Tjon, Phys. Rev. C {\bf 49}, 494 (1994); 
N. K. Devine and S. J. Wallace, Phys. Rev. C {\bf 51}, 3222 (1995).
\par\ni
\item L. Theussl and B. Desplanques, ``Crossed-boson exchange contribution and 
Bethe-Salpeter equation'', Institut des Sciences Nucleaires (Grenoble) preprint (1999)
\par\noindent
\item J. W. Darewych, M. Horbatsch and R. Koniuk, Phys. Rev. D {\bf 45}, 675 (1992).
\par\ni
\item  L. Di Leo  and J. W. Darewych,  Can. J. Phys. {\bf 71}, 365 (1993).
\par\noindent
\end{enumerate}

\vfill \eject

\ni {\bf Figure captions}
\vss \vss
\ni {\bf Figure 1} 

Two-particle bound-state energy E (eq. 39) due to virtual-annihilation interaction only, 
for various values of the coupling constant $\alpha$, at a mediating-quantum mass of 
$\mu = 2 \sqrt{2} m$ (i.e. $b = 1$). Binding sets in for $\alpha > \alpha_0 = 8.9203$, 
and  $E/m \to \epsilon_{\rm {min}} = 1.5058$ for $\mu / m = 2 \sqrt{2}$.

\vss

\ni {\bf Figure 2}

Particle-antiparticle bound state mass $E/m$ for various 
values of the coupling constant $\alpha$, 
massless quantum exchange. Curves, from highest to lowest: 
 ligh-front ladder Bethe-Salpeter [Ji and Funstahl, 20a]; 
ladder Bethe-Salpeter [Wick and Cutkosky, 17, 18];
equations (44) without virtual 
annihilation or retardation [Di Leo and Darewych, 33]; 
equation (44) with virtual 
annihilation and retardation ($n=2$);  equation (44) with retardation but without virtual 
annihilation ($n=2$); Feshbach-Villars formalism 
[Darewych, 37].  Note that all these results, save one, 
do not contain the virtual-annihilation interaction.

\vss
\ni {\bf Figure 3}

Particle-antiparticle bound state mass $E/m$ for various 
values of the coupling constant $\alpha$, 
massive quantum exchange: $\mu / m = 0.15$. Curves, 
from highest to lowest:  ladder Bethe-Salpeter 
[Wick and Cutkosky, 17, 18];  present results of equation (44) 
with virtual 
annihilation ($n=2$); equation (44) without virtual 
annihilation ($n=2$); crosses: Blankenbecler-Sugar [27b, 39]; 
  Feshbach-Villars formalism [Darewych, 37];
diamonds: Gross [27b, 40]; boxes: equal-time [27b, 41];
circles: Feynam-Schwinger formalism [27b].
  Note that all these results, save one, 
do not contain the virtual-annihilation interaction.

\vfill \eject

	%
	
	\setlength{\textwidth}{16.5cm}
	\setlength{\oddsidemargin}{0cm}
	\setlength{\evensidemargin}{0cm}

	{\bf Table 1.}

	\vspace{0.5cm}

	Ground state particle-antiparticle mass (rest energy) $E_2/m$
	obtained from eq. (50) using $f(p)=\omega_p/(p^2+b^2)^n$ for various 
	values of the coupling constant $\alpha=\frac{g^2}{16\pi m^2}$
	and variational parameter $n$. Massive exchange case, $\mu/m=0.15$. 
	Virtual annihilation is included here. The number in brackets below
	each energy is the corresponding value of $b/m$.

	\vspace{0.5cm}

	   \begin{tabular*}{14cm}{l|c@{\extracolsep{\fill}}
	      c@{\extracolsep{\fill}}c@{\extracolsep{\fill}}
		 c@{\extracolsep{\fill}} }\hline

	& $n=2$ & $n=5/2$ & $n=3$ & $n=7/2$ \\   
	$\alpha$ 
	& $E_2/m$ ($b/m$) & $E_2/m$ ($b/m$) & $E_2/m$ ($b/m$) &
	$E_2/m$ ($b/m$) \\  \hline

	0.3 & 2.0 & 2.0 & 2.0 & 2.0  \\  \hline
	    & 1.991393 & 1.991776 & 1.992553 & 1.993181 \\  
	0.5 & (0.159615) & (0.215695) & (0.255046) & (0.287139) \\  \hline
	    & 1.972650 & 1.971980 & 1.973059 & 1.974105 \\
	0.7 & (0.219966) & (0.301969) & (0.360828) & (0.409350) \\  \hline
	    & 1.946141 & 1.943654 & 1.944858 & 1.946269 \\
	0.9 & (0.267098) & (0.369155) & (0.443004) & (0.504074) \\  \hline
	    & 1.930478 & 1.926874 & 1.928095 & 1.929676 \\
	1.0 & (0.287458) & (0.398092) & (0.478350) & (0.544784) \\  \hline
	    & 1.713297 & 1.694965 & 1.695824 & 1.699076 \\
	2.0 & (0.430321) & (0.598907) & (0.722546) & (0.825342) \\  \hline
	    & 1.428109 & 1.393489 & 1.394290 & 1.399622 \\
	3.0 & (0.520198) & (0.722945) & (0.872253) & (0.996653) \\  \hline
	    & 1.104266 & 1.054041 & 1.055524 & 1.063454 \\
	4.0 & (0.585899) & (0.812489) & (0.979783) & (1.115435) \\  \hline
	    & 0.755363 & 0.690629 & 0.693494 & 0.704603 \\
	5.0 & (0.637454) & (0.882107) & (1.061143) & (1.201048) \\ \hline
	    & 0.388913 & 0.310760 & 0.315679 & 0.330580 \\
	6.0 & (0.679660) & (0.938688) & (1.122941) & (1.266242) \\  \hline

	\end{tabular*}

	\pagebreak

	{\bf Table 2.}

	\vspace{0.5cm}
	 
	Value of $E_2/m$ (ground state) for $\mu/m=0$ in $3+1$,
	eq. (50) with $n=2$.

	\vspace{0.5cm}

	   \begin{tabular*}{14cm}{l|c@{\extracolsep{\fill}}
	      c@{\extracolsep{\fill}}c@{\extracolsep{\fill}}  }\hline
	 
	   & With annihilation & Without annihilation & Di Leo [33] \\
	$\alpha$ & & & (no ann. or retard.) \\
	   & $E_2/m$ ($b/m$) & $E_2/m$ ($b/m$) & $E_2/m$ ($b/m$) \\ \hline
	& 1.9975 & 1.9975 & 1.997527 \\
	0.1 & (0.0492) & (0.0495) & (0.049049) \\ \hline
	& 1.9792 & 1.9786 & 1.979182 \\
	0.3 & (0.1327) & (0.1372) & (0.132481) \\ \hline
	& 1.9469 & 1.9441 & 1.947012 \\
	0.5 & (0.1963) & (0.2075) & (0.195480) \\ \hline
	& 1.9045 & 1.8975 & 1.904898 \\
	0.7 & (0.2463) & (0.2647) & (0.244621) \\ \hline
	& 1.8544 & 1.8413 & 1.855351 \\
	0.9 & (0.2872) & (0.3126) & (0.284642) \\ \hline
	& 1.8271 & 1.8101 & 1.828330 \\
	1.0 & (0.3053) & (0.3339) & (0.302153) \\ \hline
	& 1.4962 & 1.4203 & 1.503838 \\
	2.0 & (0.4341) & (0.4902) & (0.425538) \\ \hline
	& 1.1026 & 0.9415 & 1.121291 \\
	3.0 & (0.5156) & (0.5916) & (0.501635) \\ \hline
	& 0.6744 & 0.4120 & 0.707645 \\
	4.0 & (0.5748) & (0.6662) & (0.555879) \\ \hline
	& 0.2241 & -0.1500       & 0.274678 \\
	5.0 & (0.6208) & (0.7248) & (0.597410) \\ \hline
	\end{tabular*}

	\pagebreak

	{\bf Table 3.}
	 
	\vspace{0.5cm}
	 
	Value of $E_2/m$ (ground state) for $\mu/m=0$ in $3+1$,
	eq. (50) with $n=3$.

	\vspace{0.5cm}

	   \begin{tabular*}{14cm}{l|c@{\extracolsep{\fill}}
	      c@{\extracolsep{\fill}}c@{\extracolsep{\fill}}  }\hline

	   & With annihilation & Without annihilation & Di Leo [33] \\
	 $\alpha$ & & & (no ann. or retard.) \\
	   & $E_2/m$ ($b/m$) & $E_2/m$ ($b/m$) & $E_2/m$ ($b/m$) \\ \hline 

	& 1.9976 & 1.9976 & 1.997605 \\
	0.1         & (0.0744) & (0.0747) & (0.074211) \\ \hline
	& 1.9906 & 1.9905 & 1.990593 \\
	0.2         & (0.1443) & (0.1461) & (0.144197) \\ \hline
	& 1.9794 & 1.9790 & 1.979502 \\
	0.3         & (0.2079) & (0.2128) & (0.207603) \\ \hline
	& 1.9645 & 1.9636 & 1.964567 \\
	0.4        & (0.2647) & (0.2738) & (0.263998) \\ \hline
	& 1.9465 & 1.9445 & 1.946607 \\
	0.5          & (0.3153) & (0.3292) & (0.313968) \\ \hline
	& 1.9257 & 1.9224 & 1.925973 \\
	0.6         & (0.3606) & (0.3796) & (0.358409) \\ \hline
	& 1.9036 & 1.8974 & 1.903044 \\
	0.7         & (0.4013) & (0.4255) & (0.398195) \\ \hline
	& 1.8774 & 1.8699 & 1.878128 \\
	0.8         & (0.4383) & (0.4674) & (0.434075) \\ \hline
	& 1.8505 & 1.8403 & 1.851481 \\
	0.9         & (0.4720) & (0.5060) & (0.466668) \\ \hline
	& 1.8219 & 1.8086 & 1.823314 \\ 
	1.0         & (0.5030) & (0.5417) & (0.496469) \\ \hline

	\end{tabular*}

	\pagebreak

	{\bf Table 4.}

	\vspace{0.5cm}

	 Value of $E_2/m$ (ground state) for $\mu/m=0.15$
			   in $3+1$.  Results given in columns 2,3,4 and 6 were read off fig. 1 of [27b],
			   hence the accuracy of the last figure is questionable.

	\vspace{0.5cm}

	   \begin{tabular}{lccccccc}\hline
	 
	$\alpha$ & Niewenhuis & Equal      & Gross eq. &
	Darewych & Blankenbecler & Di Leo     & Present  results \\ 
		 & and Tjon   & Time       & (with retard.) &  
	[37]     & and Sugar  &    [33]        & ($n=2$) no annih. \\ \hline
	\\
	0.3 &      &       &       & 1.999536  &       & 2.0      & 2.0 \\
	0.4 & 1.99 &       &       & 1.99534   & 1.996 & 1.996946 & 1.996582 \\
	0.5 & 1.98 &       &       & 1.98630   & 1.989 & 1.990833 & 1.989742 \\
	0.6 & 1.96 & 1.966 & 1.969 & 1.97176   & 1.979 & 1.982148 & 1.979868 \\
	0.7 & 1.91 & 1.941 & 1.948 & 1.95081   & 1.965 & 1.971207 & 1.967249 \\
	0.8 & 1.85 & 1.907 & 1.919 & 1.92199   & 1.952 & 1.985276 & 1.952143 \\
	0.9 & 1.77 & 1.861 & 1.880 & 1.88282   & 1.933 & 1.943577 & 1.934776 \\ 
	1.0 &      &       &       & 1.82847   &       & 1.927296 & 1.915346 \\
	2.0 &      &       &       &           &       & 1.703450 & 1.637352 \\
	3.0 &      &       &       &           &       & 1.412621 & 1.262094 \\
	4.0 &      &       &       &           &       & 1.084915 & 0.830372 \\
	5.0 &      &       &       &           &       & 0.733877 & 0.361942 \\
	\\
	\hline

	\end{tabular}

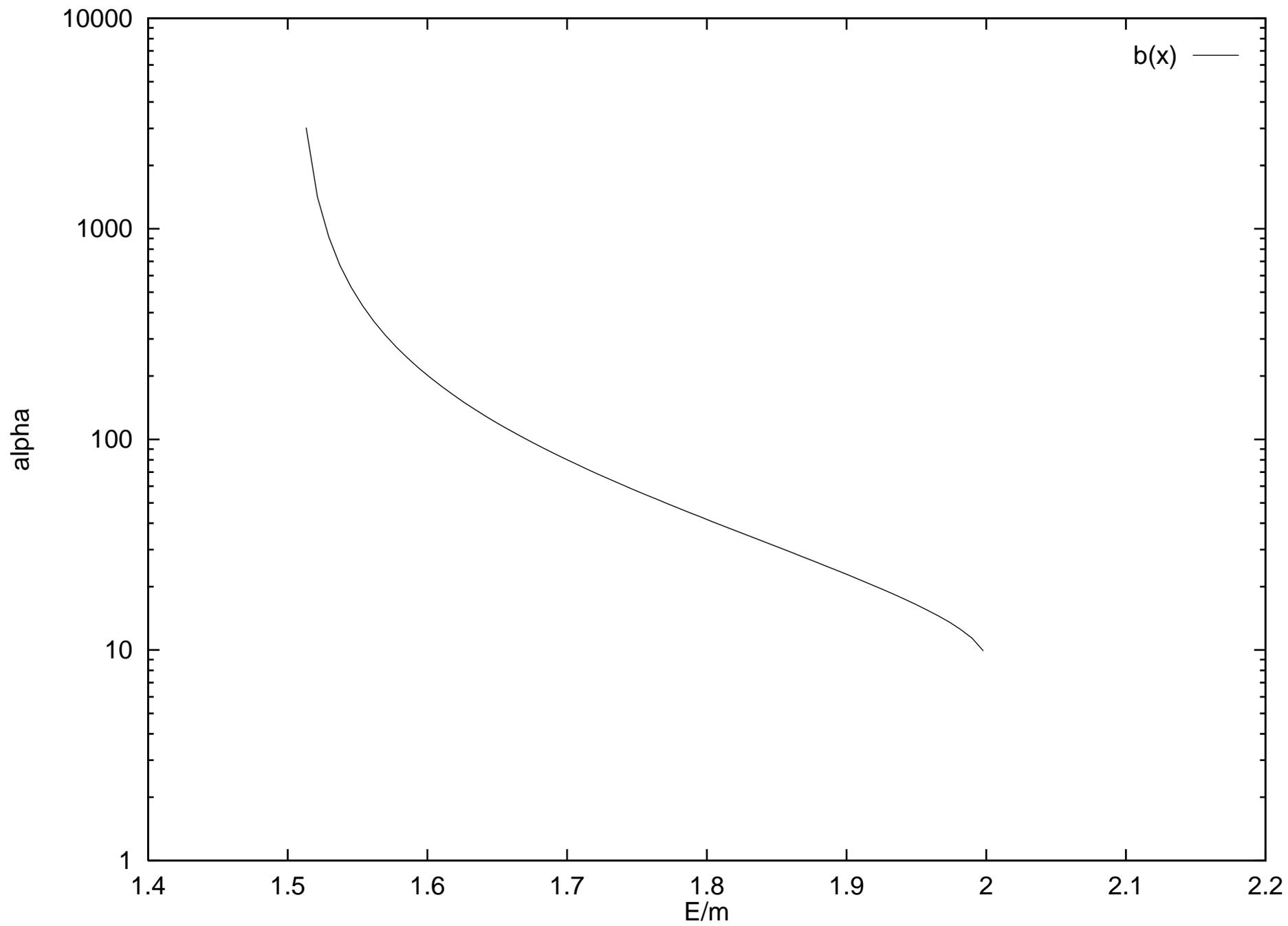
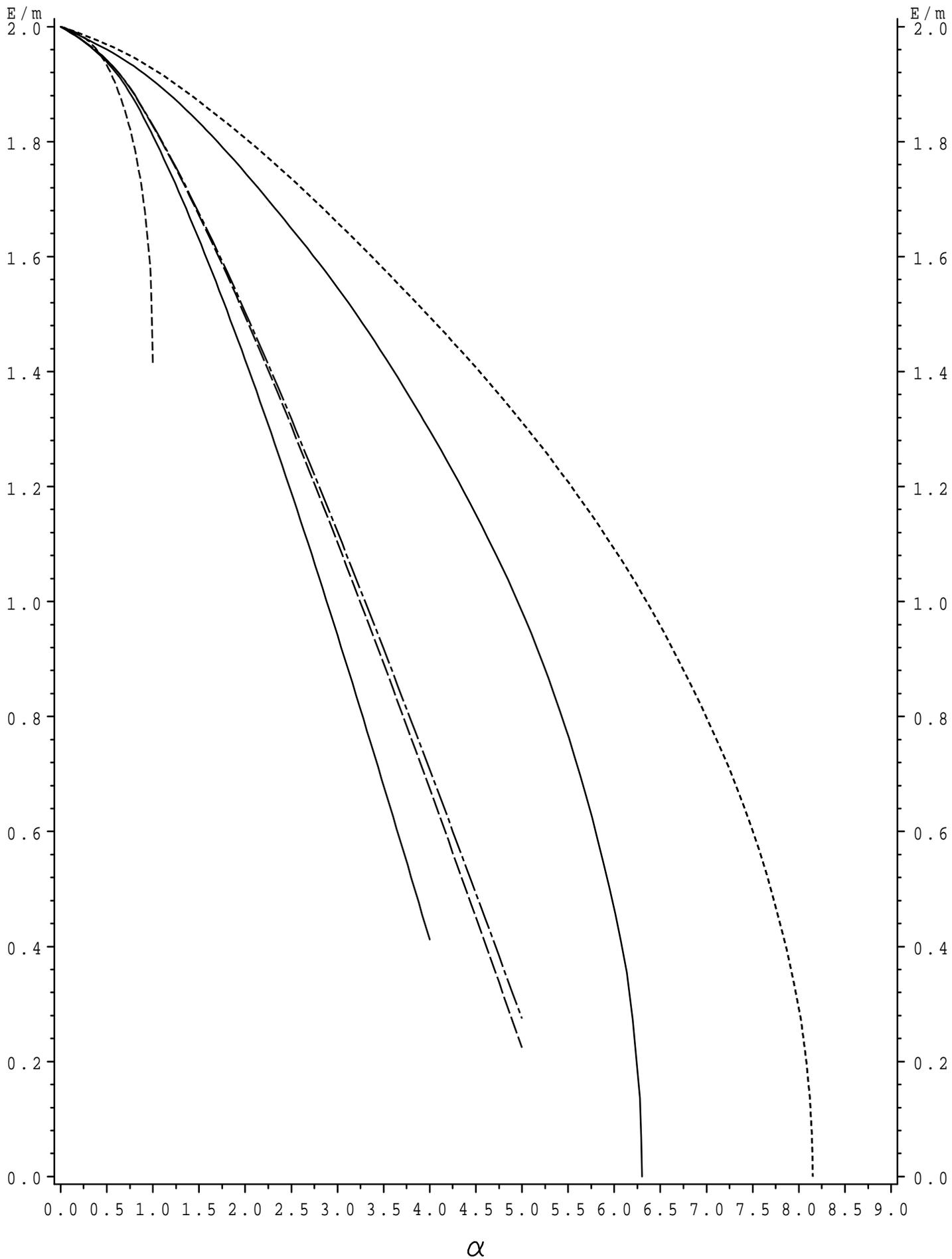
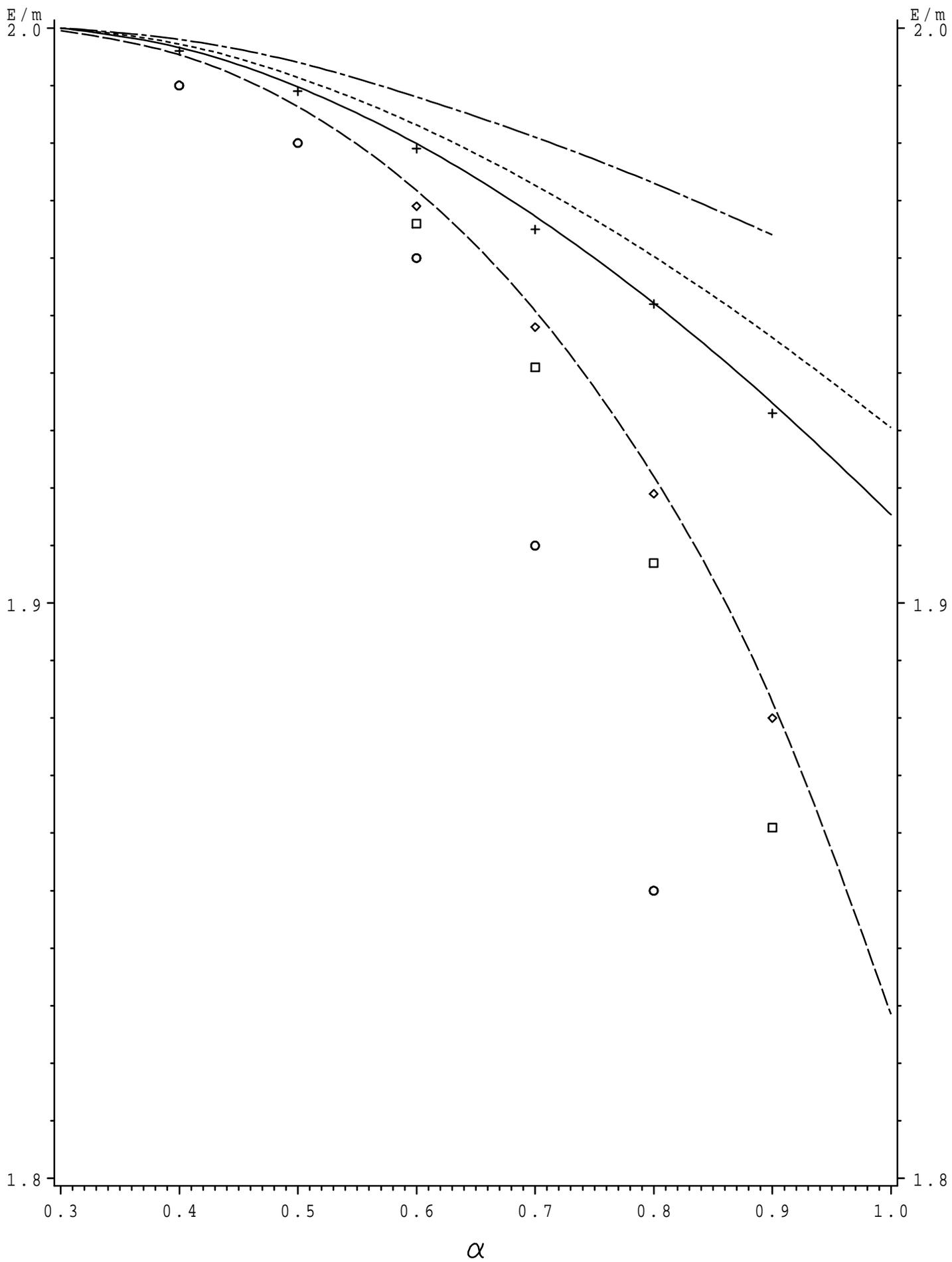
\end{document}